\documentclass[sigconf]{acmart}

\usepackage{nth}
\usepackage{stfloats}
\usepackage{multirow}
\usepackage{arydshln}
\usepackage{placeins}
\usepackage{xcolor}
\usepackage{nth}

\usepackage{enumitem}
\usepackage{todonotes}

\newcommand{\CMP}{\textsc{CMP}}
\newcommand{\NDCGb}{$\textsc{NDCG}_{bi}$}
\newcommand{\NDCGrecip}{$\textsc{NDCG}_{ic}$}
\newcommand{\NDCGlinear}{$\textsc{NDCG}_{lc}$}

\newcommand{\RDQmi}{$\text{RDQ}_{m1}$}
\newcommand{\RDQma}{$\text{RDQ}_{m2\text{-}a1}$}
\newcommand{\RDQmb}{$\text{RDQ}_{m2\text{-}a4}$}
\newcommand{\AUCt}{$\mathrm{AUC}_\tau$}

\begin{document}

\title{Rank-Deviation Quality: A Distance-Aware Metric for Multi-Answer Retrieval and Ranking Evaluation}

\author{Xiaokun Zhou}
  \affiliation{%
    \institution{Amazon}
    \country{USA}}
  \email{angzhou@amazon.com}

  \author{Alessandro Moschitti}
  \affiliation{%
    \institution{Amazon}
    \country{USA}}
  \email{amosch@amazon.com}

  \author{Danielle Class}
  \affiliation{%
    \institution{Amazon}
    \country{USA}}
  \email{dcclss@amazon.com}

\begin{abstract}


We introduce Rank-Deviation Quality (RDQ), an evaluation metric for retrieval and ranking systems that adapts to queries with varying numbers of reference items, from a single correct answer to many valid results. RDQ scores a candidate ranking against an ordered reference list (ORL): each retrieved reference item contributes its output-position weight multiplied by a rank-deviation penalty, and items outside the ORL receive zero credit. Application-specific parameters control tolerance to misordering. Larger values emphasize retrieving valid reference items, whereas smaller values place more weight on matching their reference order. The output-position weights can reflect visibility in the application's interface, such as a vertical list or a carousel. Unlike metrics that require absolute relevance grades, RDQ operates on ordinal rankings, which annotators can produce through pairwise or listwise judgments. Unlike rank-correlation measures such as Kendall's~$\tau$, RDQ accounts for both which items are returned and how they are ordered. On a 5,000-query point-of-interest (POI) dataset with 12 systems, RDQ has the highest median empirical power@100 among the 13 evaluated metric configurations. It reaches mean $\tau \geq 0.8$ agreement with its own full-query ordering at 200 queries; RBP(0.9), the strongest tested non-RDQ configuration, reaches the same threshold at 250. On TREC Deep Learning benchmarks, where NDCG uses native graded labels and RDQ uses ordinal tiers derived from them, RDQ reaches comparable median power at $n=25$, while NDCG is higher at $n=100$.

\end{abstract}

\maketitle

\section{Introduction}

Information retrieval systems are commonly evaluated with metrics such as NDCG~\cite{jarvelin2002ndcg}, RBP~\cite{moffat2008rbp}, and MAP~\cite{buckley2000evaluating}, which reward placing relevant results near the top of a ranked list. Many applications rank candidate sets of points of interest, products, videos, entities, or knowledge cards to determine which items appear near the top. When several items are acceptable, quality depends both on which items are returned and on their relative order, often without a well-defined absolute relevance scale.

Existing metric families leave a gap in this setting. Binary metrics cannot distinguish whether better reference items are placed above worse ones. Graded metrics such as NDCG and ERR~\cite{jarvelin2002ndcg,chapelle2009expected} can express such differences but require labels on a fixed relevance scale. Standard rank correlations such as Kendall's~$\tau$ and Spearman's~$\rho$ compare orderings directly but require both rankings to cover the same item set. Restricting them to the overlap ignores retrieval errors: a system can return only low-ranked reference items and still receive perfect correlation if their relative order is preserved.

Fixed-scale relevance labels are difficult to calibrate across diverse queries~\cite{voorhees1998variations,sormunen2002liberal,scholer2013effect}. The meaning of a grade for a query with one answer may differ from its meaning for a broad query with many acceptable results. Ordinal supervision instead specifies relative preferences among items or tiers without requiring globally calibrated grades~\cite{carterette2008here,roitero2022budget}.

We introduce \emph{Rank-Deviation Quality (RDQ)} for multi-answer retrieval and ranking evaluation. RDQ scores a candidate ranking against an ordered reference list (ORL): each retrieved reference item contributes its output-position weight multiplied by a rank-deviation penalty, and items outside the ORL receive zero credit. Application-specific tolerance parameters control the misordering penalty: larger values penalize misordering less, so the score rewards mainly including valid reference items, while smaller values penalize misordering more, rewarding their exact reference order. RDQ also supports tied tiers, query-varying ORL sizes, and output-position weights for different user-interface locations.

We evaluate RDQ through controlled examples and two empirical settings: 12 systems on a 5,000-query point-of-interest (POI) dataset and four systems on pooled TREC Deep Learning 2019–2022 queries. The analyses measure discriminative power, system-ranking stability, reference-list-size effects, parameter sensitivity, and behavior when graded judgments are converted to ordinal tiers.

Our contributions are as follows:
\begin{itemize}[nosep, leftmargin=*]

    \item We propose RDQ, an effectiveness metric that evaluates multi-answer retrieval and candidate rankings against query-specific ordered reference lists (ORLs), rather than absolute relevance grades.

    \item We define RDQ using two rank-deviation penalties, with support for tied tiers, variable ORL sizes, application-specific misordering tolerance, and output-position weights for user-interface locations.
    
    \item We evaluate RDQ on 12 systems over 5,000 POI queries. \RDQma{} has the highest median empirical power@100 among 13 configurations and reaches mean Kendall's $\tau \geq 0.8$ with 200 queries, compared with 250 for RBP(0.9), the strongest tested non-RDQ configuration. Sixteen of the 25 $M_2$ configurations also have higher median power@100 than RBP(0.9).

    \item We evaluate RDQ on the public TREC-DL 2019--2022 datasets using ordinal tiers derived from native graded judgments. \RDQmi{} and native NDCG have comparable median power at $n=25$ (0.108 versus 0.102), while NDCG is higher at $n=100$ (0.437 versus 0.367). Across the four-system panel, per-year correlations show broad agreement with established metrics in 2021--2022 and weaker agreement in 2020.
    
\end{itemize}

The remainder of the paper is organized as follows. Section~\ref{sec:related-work} reviews related work, and Section~\ref{sec:def} defines RDQ. Section~\ref{sec:experimental-analysis} presents the experiments, discussion, and limitations. Section~\ref{sec:conclusion} concludes the paper.

\section{Related Work}\label{sec:related-work}
Evaluation metrics for ranked retrieval differ in their judgment inputs, user models, and treatment of ranking errors.

\vspace{-.3em}
\paragraph{Gain-based effectiveness metrics.}
Average Precision averages precision at ranks containing relevant items, and
MAP averages this quantity across queries~\cite{manning2008ir}. It therefore
uses binary relevance and does not distinguish quality differences among
relevant items. NDCG instead discounts graded gains by output position and
normalizes by the ideal ranking~\cite{jarvelin2002ndcg}. ERR maps graded labels
to stopping probabilities in a cascade model~\cite{chapelle2009expected}.
These metrics can represent quality levels, but require a numeric gain or
satisfaction mapping, whose spacing is not determined by tier order alone.
The underlying judgments can also vary across assessors and assessment
contexts~\cite{voorhees1998variations,scholer2013effect}. RDQ instead consumes
ordinal tiers directly; our experiments separately test two rank-to-grade conversions of the same ORL for NDCG.

\vspace{-.3em}
\paragraph{Ranking and rank-biased comparisons.}
Standard Spearman and Kendall coefficients compare complete rankings over the
same item set~\cite{carterette2009rankcorrelation} using unweighted item or pair
comparisons. Extensions compare partial or top-$k$ rankings~\cite{fagin2003topk}
and introduce element and position weights~\cite{kumar2010generalized}. RBO
compares indefinite rankings that need not share the same items, through
geometrically weighted prefix overlap~\cite{webber2010rbo}. Tie-aware variants formalize prefix evaluation for tied rankings~\cite{corsi2024ties}. 

RBP is a top-weighted effectiveness metric based on a geometric persistence model~\cite{moffat2008rbp}. Recent rank-biased measures cover additional combinations of set and ranking inputs~\cite{Moffat2024RBR}. Rank-Biased Recall compares an observation set with a reference ranking, ignoring order within the observation. Rank-Biased Alignment rewards shared items that are jointly high in both rankings, using the average of their two ranks. RDQ instead separates output-position importance from the penalty for deviation from the expected ORL rank.

\vspace{-.3em}
\paragraph{Preference-based evaluation.}
Pairwise preferences can be easier to elicit than absolute grades and can supply an ordinal reference~\cite{carterette2008here,roitero2022budget}.
Preference-based retrieval evaluation is an established line: Yao models
effectiveness from document preferences~\cite{yao1995preference}, and subsequent
work defines measures directly over preference judgments~\cite{sakai2020preferences}. Clarke et al. define \emph{compatibility} (\CMP{}) as the maximum RBO similarity between a system ranking and ideal rankings consistent with a query-specific weak order~\cite{clarke2020withoutgain,clarke2020maximum}. \CMP{} supports tied tiers, a query-varying number of tiers, and system outputs that need not contain the same items as the reference, and it requires no numeric gain values. RDQ addresses a similar ordinal-reference setting but scores each retrieved reference item by its deviation from a target rank rather
than by prefix overlap.

\vspace{-.3em}
\paragraph{Evaluating evaluation metrics.}

We use established meta-evaluation tools. Randomization, bootstrap, and paired
$t$-tests have shown similar reliability for mean IR comparisons, and we use
two-sided randomization tests~\cite{smucker2007comparison}. We estimate power
and required query counts following work on IR power and topic-set-size
design~\cite{webber2008power,sakai2016topicsetsize}, and report the proportion
of system pairs distinguished by each metric~\cite{sakai2006bootstrap}.
Finally, we assess system-ranking stability by subsampling queries and comparing
each induced ranking with the full-query ranking using Kendall's
$\tau$~\cite{voorhees2002stability,urbano2013reliability}. Section~\ref{sec:experimental-analysis} applies identical protocols to RDQ and all evaluated baselines.

\section{Metric Definition}\label{sec:def}

This section defines RDQ in three steps. First, we specify the supplied \emph{Ordered Reference List (ORL)} and its item-to-rank mapping. Second, we define the RDQ score as a normalized weighted sum over retrieved reference items, where credit depends on output-position weights and rank-deviation penalties. Third, we instantiate the penalty function with two alternatives: a simple asymmetric penalty $M_1$ and a balanced distance-aware penalty $M_2$.

\vspace{-.3em}
\subsection{Ordered Reference Lists and ORL Ranks}\label{sec:orl_rank}

For each query $q$, an ORL is a nonempty mapping that assigns each reference
item an integer \emph{ORL rank}, where a smaller rank denotes a better item and
items sharing a rank form a tier. Ranks start at 1 and are consecutive with no
gaps, so an ORL with $T_q$ tiers uses exactly the ranks $1$ through $T_q$. Here, ``list'' means the ordering of tiers rather than a sequence with one item at each position. Items within the same tier are tied, so their order does not affect RDQ.

The ORL can be supplied in either of two equivalent forms. When no items are
tied, it is given as a ranked list, and each item takes its one-based position
as its rank: $[d_1,d_2,d_3]$ becomes $\{d_1{:}1,d_2{:}2,d_3{:}3\}$. When items
are tied, it is given directly as the item-to-ORL-rank mapping, with tied items
sharing a rank: in $\{d_1{:}1,d_3{:}1,d_2{:}2\}$, $d_1$ and $d_3$ share the
first tier.

RDQ assigns zero credit to items outside the ORL and therefore assumes that the ORL is sufficiently complete for evaluation at cutoff $k$. ORL length is
the number of items in the list and varies by query based on the query breadth. For example, a query for a particular book edition may have one valid answer, whereas a query for ``science-fiction novels'' may have many. A one-item ORL should therefore represent a narrow query, not incomplete annotation of a broad one.

RDQ takes the ORL as input and does not prescribe how it is produced. Construction reuses established annotation and preference-aggregation methods
and is orthogonal to the metric itself. Example routes include:
\begin{enumerate}[nosep, leftmargin=*]
  \item \emph{Listwise annotation.} Annotators, or an authoritative source, produce either a ranked list or an ordered list of tiers, used directly as the ORL.
  \item \emph{Pairwise-preference conversion.} Annotators indicate which of two items is better; such judgments have been used to build top-$k$ preference references for offline evaluation~\cite{clarke2021topk}. Established methods can aggregate the comparisons into a ranking, including the Bradley--Terry model~\cite{bradley1952rank}, Rank Centrality~\cite{negahban2017rankcentrality}, and crowdsourcing-aware variants~\cite{chen2013pairwise}). They can also use  sparse comparisons and prioritize informative pairs under a fixed budget~\cite{roitero2022budget}. Items whose aggregated scores are not significantly separated share an ORL rank.
  \item \emph{Graded-relevance conversion.} Each distinct positive grade becomes a tier. Items are processed by decreasing grade, and each item is assigned the consecutive ORL rank of its grade, starting at 1. Section~\ref{sec:ir-benchmarks} demonstrates this conversion on public TREC labels.
\end{enumerate}

\textbf{Why ORL ranks rather than graded gains.}
An ORL records relative preferences among reference items and tiers. When it is constructed directly from listwise or pairwise judgments, assessors need not assign every item to a fixed relevance category. Applying such category boundaries consistently can be difficult: assessors disagree even on binary relevance~\cite{voorhees1998variations}, and grading thresholds can drift with the documents encountered during judging~\cite{scholer2013effect}. Moreover, roughly half of the items judged relevant under binary criteria were only marginally relevant when re-assessed on a graded scale ~\cite{sormunen2002liberal}. Graded measures must also map levels to gain values, and these mappings are typically heuristic and fixed in advance ~\cite{jarvelin2002ndcg}. RDQ instead operates directly on ORL ranks and requires no gain mapping.

\vspace{-.3em}
\subsection{Rank-Deviation Quality (RDQ) metric}\label{sec:RDQ}

RDQ takes two inputs per query: the reference list (the ORL) and a candidate
list (a system's ranked output truncated at cutoff $k$). Only items that appear
in the ORL earn credit; items outside the ORL contribute nothing. For each
retrieved reference item, RDQ combines three factors: (1) an output-position
weight that reflects the importance of the displayed position, (2) a
rank-deviation penalty that measures how far the item's ORL rank is from the
rank expected at its position, and (3) a normalization term that makes the
score equal to $1$ for the ideal ranking.

Let $R_q=[r_1,\ldots,r_m]$ denote the $m$ reference items of the ORL, and let
$\gamma(x)$ denote the ORL rank of item $x$ under the mapping of
Section~\ref{sec:orl_rank}, with smaller values indicating better items and
tied items sharing a value. Items not appearing
in $R_q$ receive zero credit under this evaluation. Let
$P_q@k=[p_1,\ldots,p_k]$ be the system output truncated at rank $k$. If the system returns fewer than $k$ items, the empty positions contribute nothing to the numerator while the denominator is unchanged. If the same item appears multiple times, only its first occurrence is eligible for credit. Later occurrences contribute zero because repeated display of the same item should not increase retrieval quality. We define
$N_k=\min(k,|R_q|)$, the number of positions an ideal ranking can fill with
reference items, where $|R_q|$ is the number of reference items for query
$q$.

The RDQ score is:
\vspace{-.3em}

\begin{equation}
  \mathrm{RDQ}@k=
  \frac{
  \sum_{i=1}^{k}
  \mathbf{1}\left[
  p_i \in R_q
  \land
  p_i \notin \{p_1,\ldots,p_{i-1}\}
  \right]
  \,\mu_i\,
  \rho\left(i,\gamma_i^\star,\gamma(p_i)\right)
  }{
  \sum_{i=1}^{N_k}\mu_i
  },
  \label{eq:rdq}
  \end{equation}

where:
\begin{itemize}[nosep, leftmargin=*]
  \item $\mu_i$ is the output-position weight for displayed position $i$. They are positive and non-increasing, $\mu_1 \geq \mu_2 \geq \cdots >0$, so a later displayed position is never assigned more importance than an earlier one. The weights depend only on the layout, not on the number of reference items $|R_q|$. The weights reflect user-interface importance and can be adapted to the application. If the application does not distinguish positions, all weights can be set equal (e.g., $\mu_i=1$), so every displayed position contributes equally. For a vertical layout, one may instead use $\mu_1=1$, $\mu_2=\mu_3=0.8$, $\mu_4=\ldots=\mu_{10}=0.5$, and $\mu_i=0.1$ for all remaining positions. A carousel with three initially visible items may use $\mu_1=\mu_2=\mu_3=1$ and $\mu_i=0.1$ for all remaining positions. These weights may also be learned from interaction models.
  \item $\gamma_i^\star$ is the ideal ORL rank at position $i$, determined by the tier sizes alone. The order of items within a tier does not matter. Without ties, $\gamma_i^\star=i$, where $i$ starts from 1. With ties, all positions within a tier share that tier's ORL rank. For example, if $R_q=[r_1,r_2,r_3]$ has ORL ranks $[1,2,2]$, then $\gamma_1^\star=1$, $\gamma_2^\star=2$, and $\gamma_3^\star=2$, because the second tier occupies positions 2 and 3. For positions beyond the ORL ($i>|R_q|$), $\gamma_i^\star$ continues with the next consecutive ranks (here $\gamma_4^\star=3$ and $\gamma_5^\star=4$), so a reference item placed below the ORL is still charged its displacement.
  \item $\gamma(p_i)$ is the ORL rank of the item at candidate position $i$. If $p_i$ matches a reference item, $\gamma(p_i)$ equals that item's ORL rank. Otherwise, $p_i$ is not in the ORL and receives zero credit. Using the same $R_q$, suppose the system output is $[p_1,p_2,p_3]$ where $p_1=r_2$, $p_2=r_3$, and $p_3=r_1$. Then $\gamma(p_1)=2$ because $p_1$ is $r_2$. Similarly, $\gamma(p_2)=2$ because $r_3$ shares rank 2, and $\gamma(p_3)=1$ because $p_3$ is $r_1$.
  \item $\rho(i,\gamma_i^\star,\gamma(p_i))\in[0,1]$ is the penalty for placing item $p_i$ at candidate position $i$, defined in Section~\ref{sec:penalty}.
  \item $\mathbf{1}\left[p_i \in R_q \land p_i \notin {p_1,\ldots,p_{i-1}}\right]$ equals 1 when the retrieved item appears in the ORL and has not occurred at an earlier output position, and equals 0 otherwise. For the first candidate position, the set of preceding items is empty.
\end{itemize}

Under the first-occurrence rule and non-increasing position weights, at most $N_k$ distinct ORL items can contribute, each with penalty at most 1. The numerator is therefore at most $\sum_{i=1}^{N_k}\mu_i$,
which is the maximum achievable score at cutoff $k$. Consequently, $0\leq\mathrm{RDQ}@k\leq1$, and an ideal ranking attains $\mathrm{RDQ}@k=1$.

\vspace{-.3em}
\subsection{Penalty Weighting Methods}\label{sec:penalty}

The penalty function $\rho$ determines how much credit a retrieved reference
item receives when it appears at a position whose expected ORL rank differs
from the item's ORL rank. A value of $\rho=1$ indicates no penalty, while
smaller values indicate more severe rank deviation. Items not
appearing in the ORL receive zero credit, so $\rho$ is defined only for
retrieved reference items. We consider two penalty functions:
a simple asymmetric penalty ($M_1$) and a balanced distance-aware penalty
($M_2$).

\vspace{-.3em}
\subsubsection{\textbf{Method \(M_1\): Simple Asymmetric Penalty}}
The first penalty assigns credit according to the ratio between the expected
ORL rank at position $i$ and the ORL rank of the retrieved item:

\vspace{-.3em}
\begin{equation}
\rho_{M_1}=
\begin{cases}
\min\left(1,\frac{\gamma_i^\star}{\gamma(p_i)}\right), & p_i\in R_q,\\
0, & \text{otherwise}.
\end{cases}
\label{eq:m1}
\end{equation}

This penalty penalizes placing a lower-quality reference item above a
higher-quality one. However, it is asymmetric. If a highly ranked reference
item is retrieved below its ideal position, $M_1$ may assign no penalty as
long as the retrieved item has a better ORL rank than the item expected at
that position.

\vspace{-.3em}
\subsubsection{\textbf{Method $M_2$: Balanced Distance-Aware Penalty}}
The second penalty uses absolute rank displacement and maps it through an exponential whose decay rate is set by an application-tunable tolerance.

We define
the rank deviation $d(i,p_i)$ as:
\vspace{-.3em}
\begin{equation}
d(i,p_i)=|\gamma_i^\star-\gamma(p_i)|.
\label{eq:deviation}
\end{equation}
The $M_2$ penalty is then:
\begin{equation}
\rho_{M_2}=
\begin{cases}
\exp\left(
\frac{-d(i,p_i)}
{\sqrt{|R_q|}\cdot\alpha\cdot(1+\lambda(\gamma(p_i)-1))}
\right), & p_i\in R_q,\\
0, & \text{otherwise}.
\end{cases}
\label{eq:m2}
\end{equation}

\noindent where $|R_q|$ is the number of reference items for query $q$;
$\alpha>0$ sets global tolerance for displacement, with larger values
weakening the penalty; and $\lambda\geq0$ controls how tolerance changes with
reference rank. At $\lambda=0$, tolerance is independent of reference rank.
Higher values weaken penalties more for lower-ranked reference items. We
treat $\alpha$ and $\lambda$ as application-specific tolerance parameters.
Unless stated otherwise, we fix $\lambda=0.2$ and report two settings of $\alpha$: $\alpha=1$ (a stronger displacement penalty) and $\alpha=4$ (a weaker one).
Section~\ref{sec:sensitivity} reports median power over 25
configurations rather than selecting a configuration from the evaluation
outcome.

$M_2$ has three intended properties. First, larger deviations from the target
ORL rank receive smaller credit because the exponential decreases with
$d(i,p_i)$. Second, for the same deviation, the factor
$1+\lambda(\gamma(p_i)-1)$ penalizes a better-ranked reference item more
strongly than a lower-ranked one. Third, $\sqrt{|R_q|}$ makes a fixed displacement more costly for a short ORL than for a long one. Because displacement is measured in ORL-tier units while $|R_q|$ counts items, larger tied tiers also weaken the penalty. Section 4.5 discusses this behavior on TREC-DL. The exponential is steepest near the origin. As normalized deviation grows, credit approaches
zero and the marginal effect of additional displacement diminishes.

\section{Experimental Analysis}\label{sec:experimental-analysis}
We evaluate RDQ in three settings. Controlled examples that isolate the metric's behavior (Section~\ref{sec:controlled}). The POI experiments measure discriminative power, reference-list-size effects, ranking stability, and parameter sensitivity across 12 reranking systems (Section~\ref{sec:disc-power}). The TREC Deep Learning experiments evaluate RDQ on tie-heavy ORLs derived from graded relevance judgments (Section~\ref{sec:ir-benchmarks}).

\vspace{-.3em}
\subsection{Common Experimental Setup}\label{sec:poi-common-setup}
Across these settings, we compare three RDQ configurations. \RDQmi{} uses $M_1$. \RDQma{} and \RDQmb{} use $M_2$ with $\alpha=1$ and $\alpha=4$, respectively; the former applies the stronger displacement penalty. For $M_2$, we use $\lambda=0.2$ unless otherwise specified. All variants use the default output-position weights for a vertical search layout\footnote{\label{fn:ui-weight} The default output-position weights are [1, 0.9, 0.8, 0.8, 0.7, 0.6, 0.5, 0.5, 0.4, 0.4, 0.4, 0.3, 0.3,  0.2, 0.2, 0.2], with weight \(0.1\) for remaining positions.}.

Binary NDCG (\NDCGb), MAP, RBP, and \CMP{} are evaluated in all analyses.  The controlled examples in  Section~\ref{sec:controlled} use RBP(0.8) and CMP(0.95). The POI and TREC-DL analyses evaluate multiple persistence settings for RBP and CMP in Section~\ref{sec:internal} and Section~\ref{sec:ir-benchmarks}.

The graded NDCG variants use the exponential gain $2^{\mathrm{grade}}-1$, and they differ only in where the grades come from. On TREC-DL (Section~\ref{sec:ir-benchmarks}) the grades are the native $0$--$3$ judgments, so no conversion is needed. In Sections~\ref{sec:controlled} and~\ref{sec:internal}, which have no native grades, we convert each ORL rank $r$ to a grade in two ways: \textit{inverse conversion} (\NDCGrecip) uses $\mathrm{grade}(r)=1/r$, and \textit{linear conversion} (\NDCGlinear) uses $\mathrm{grade}(r)=T_q-r+1$ (best tier highest), where $T_q$ is the number of tiers as defined in Section~\ref{sec:orl_rank} (e.g., $T_q=2$ for the ORL $[1,2,2]$). Spearman's $\rho$ and Kendall's $\tau$ appear only in the controlled examples (Section~\ref{sec:controlled}) as rank-correlation references when applicable. We omit them when the item sets differ or either rank vector is constant, since comparing only shared items would ignore missing and non-reference items.

All metrics use cutoff $k=5$ in Sections~\ref{sec:controlled} and ~\ref{sec:internal}. Section~\ref{sec:ir-benchmarks} uses $k=10$, following the standard TREC-DL reporting depth.

\vspace{-.3em}
\subsection{Analysis with Controlled Examples}\label{sec:controlled}

We first use controlled examples to illustrate the behavior defined in Section~\ref{sec:def}. Each block of Table~\ref{tab:controlled} isolates one property.

\begin{table*}[t]
    \centering
    \small
    \setlength{\tabcolsep}{2pt}
    \caption{Controlled examples. ORLs are shown as $\{\text{item}{:}\text{rank}\}$ mappings. Each block isolates one property. $0$ denotes an item outside the ORL. Tau and Rho are omitted (--) when the item sets differ or either rank vector is constant. CMP uses $p=0.95$. Exp10 uses top-focused output-position weights $\mu^{\star}=[1, 0.2, 0.2, 0.1, 0.1]$ instead of the default weights.}
    \vspace{-.8em}
    \begin{tabular}{l l l c c c c c c c c c c c}
    \toprule

    \textbf{Exp} &
    \textbf{\shortstack{ORL rank\\mapping}} &
    \textbf{Prediction}(k=5) &
    \footnotesize{MAP} &
    \footnotesize{Tau} &
    \footnotesize{Rho} &
    \footnotesize{RBP(0.8)} &
    \footnotesize{\NDCGb} &
    \footnotesize{\NDCGrecip} &
    \footnotesize{\NDCGlinear} &
    \footnotesize{\CMP} &
    \footnotesize{\RDQmi} &
    \footnotesize{\RDQma} &
    \footnotesize{\RDQmb} \\

    \midrule
    \multicolumn{14}{l}{\emph{Positional sensitivity}} \\

    Exp1 &
    \multirow{3}{*}{\shortstack[l]{$\{d_1{:}1,d_2{:}2,d_3{:}3,$\\$d_4{:}4,d_5{:}5\}$}} &
    {[}$d_1$,$d_2$,$d_3$,$d_4$,$d_5${]}
     & 1.0000 & 1.0000 & 1.0000 & 0.6723 & 1.0000 & 1.0000 & 1.0000 & 1.0000 & 1.0000 & 1.0000 & 1.0000 \\
    Exp2 &
     &
    {[}$d_2$,$d_1$,$d_3$,$d_4$,$d_5${]}
     & 1.0000 & 0.8000 & 0.9000 & 0.6723 & 1.0000 & 0.8587 & 0.8706 & 0.7790 & 0.8810 & 0.8487 & 0.9562 \\
    Exp3 &
     &
    {[}$d_3$,$d_1$,$d_2$,$d_4$,$d_5${]}
     & 1.0000 & 0.6000 & 0.7000 & 0.6723 & 1.0000 & 0.8083 & 0.7830 & 0.6740 & 0.8413 & 0.7511 & 0.9252 \\
    \midrule
    \multicolumn{14}{l}{\emph{Severe misranking and invalid items}} \\

    Exp4 &
    \multirow{2}{*}{$\{d_1{:}1,\ldots,d_{20}{:}20\}$} &
    {[}$d_1$,$d_2$,$d_3$,$d_{4}$,$d_{20}${]}
     & 0.2500 & -- & -- & 0.6723 & 1.0000 & 0.9713 & 0.9841 & 0.9640 & 0.8750 & 0.9162 & 0.9733 \\
    Exp5 &
     &
    {[}$d_1$,$d_2$,$d_3$,$d_4$,0{]}
    & 0.2000 & -- & -- & 0.5904 & 0.8688 & 0.9624 & 0.9841 & 0.9640 & 0.8333 & 0.8333 & 0.8333 \\
    \midrule
    \multicolumn{14}{l}{\emph{Reference-list size adaptation}} \\

    Exp6 &
    $\{d_1{:}1\}$ &
    {[}0,$d_1$,0,0,0{]}
     & 0.5000 & -- & -- & 0.1600 & 0.6309 & 0.6309 & 0.6309 & 0.5355 & 0.9000 & 0.3311 & 0.7009 \\
     \noalign{\vspace{2pt}}
     \hdashline[1pt/4pt]
     \noalign{\vspace{2pt}}
    Exp7 &
    $\{d_1{:}1,\ldots,d_6{:}6\}$ &
    {[}0,$d_1$,$d_2$,$d_3$,$d_4${]}
     & 0.4528 & -- & -- & 0.4723 & 0.6608 & 0.6686 & 0.6836 & 0.5241 & 0.7619 & 0.5494 & 0.7019 \\
    \noalign{\vspace{2pt}}
    \hdashline[1pt/4pt]
    \noalign{\vspace{2pt}}
    Exp8 &
    $\{d_1{:}1,\ldots,d_{20}{:}20\}$ &
    {[}0,$d_1$,$d_2$,$d_3$,$d_4${]}
     & 0.1358 & -- & -- & 0.4723 & 0.6608 & 0.6686 & 0.6829 & 0.5241 & 0.7619 & 0.6367 & 0.7284 \\
    \midrule
    \multicolumn{14}{l}{\emph{Output-position importance}} \\

    Exp9 &
    \multirow{2}{*}{\shortstack[l]{$\{d_1{:}1,d_2{:}2,$\\$d_3{:}3,d_4{:}4\}$}} &
    {[}$d_1$,0,$d_3$,$d_4${]}, \footnotesize{default $\mu$}
    & 0.6042 & -- & -- & 0.4304 & 0.7537 & 0.8226 & 0.7931 & 0.7356 & 0.7429 & 0.7429 & 0.7429 \\
    Exp10 &
     &
    {[}$d_1$,0,$d_3$,$d_4${]}, $\mu^{\star}$
    & 0.6042 & -- & -- & 0.4304 & 0.7537 & 0.8226 & 0.7931 & 0.7356 & 0.8667 & 0.8667 & 0.8667 \\
    \midrule
    \multicolumn{14}{l}{\emph{Reversed order and tied relevance}} \\

    Exp11 &
    \multirow{2}{*}{$\{d_1{:}1,d_2{:}2\}$} &
    {[}$d_1$,$d_2${]}
     & 1.0000 & 1.0000 & 1.0000 & 0.3600 & 1.0000 & 1.0000 & 1.0000 & 1.0000 & 1.0000 & 1.0000 & 1.0000 \\
    Exp12 &
     &
    {[}$d_2$,$d_1${]}
    & 1.0000 & -1.0000 & -1.0000 & 0.3600 & 1.0000 & 0.8286 & 0.7967 & 0.6975 & 0.7368 & 0.5255 & 0.8512 \\
    \noalign{\vspace{2pt}}
    \hdashline[1pt/4pt]
    \noalign{\vspace{2pt}}
    Exp13 &
    $\{d_1{:}1,d_2{:}1\}$ &
    {[}$d_2$,$d_1${]}
    & 1.0000 & -- & -- & 0.3600 & 1.0000 & 1.0000 & 1.0000 & 1.0000 & 1.0000 & 1.0000 & 1.0000 \\
    \bottomrule
    \end{tabular}
    \label{tab:controlled}
  \vspace{-.3em}
\end{table*}

\emph{Positional sensitivity (Exp1--3).}\label{ex:ce1}
Here the ORL is a strict list of five distinct items (ranks 1--5). Exp1 is the ideal prediction, Exp2 swaps the top two items, and Exp3 promotes $d_3$ to the first position. MAP, RBP, and \NDCGb{} assign identical scores because every retrieved item is in the ORL. Spearman's $\rho$ and Kendall's $\tau$, \NDCGrecip{}, \NDCGlinear{}, \CMP{}, and all RDQ variants decrease from Exp1 to Exp3, so positional sensitivity is shared among the order-aware metrics. \RDQmb{} shows the smallest drop because $\alpha{=}4$ weakens the rank-deviation penalty.

\emph{Severe misranking and invalid items (Exp4--5).}\label{ex:ce}\label{ex:ce5}
Here the ORL contains 20 distinct items (ranks 1--20). Exp4 returns the top four reference items followed by $d_{20}$; Exp5 replaces $d_{20}$ with an item outside the ORL. Standard $\tau$ and $\rho$ are omitted because the candidates and ORL contain different item sets. In Exp4, RBP (0.6723) and \NDCGb{} (1.0000) equal their ideal top-five values, ignoring the severe displacement of $d_{20}$. Both drop in Exp5 only because a relevant item is lost. 
At the reported precision, NDCG${lc}$ and CMP have the same value in both rows: 0.9841 and 0.9640, respectively. Neither visibly separates the displaced reference item from the non-ORL item. NDCG${ic}$ changes slightly, from 0.9713 to 0.9624.
RDQ instead gives the displaced item partial credit in Exp4 and zero in Exp5: the weaker-penalty \RDQmb{} changes most (0.9733 to 0.8333), and all three RDQ variants coincide at 0.8333 in Exp5 because the non-ORL position contributes zero regardless of penalty.

\emph{Reference-list size adaptation (Exp6--8).}\label{ex:ce2}
Here the three ORLs have 1, 6, and 20 items. Each prediction places an item outside the ORL at position 1 and shifts the retrieved reference items down. The \(M_2\) scores increase with ORL size, while RBP, \NDCGb{}, and \NDCGrecip{} rise from Exp6 to Exp7 and then plateau, and \CMP{} stays nearly flat across the three ORLs. This behavior reflects the intended size normalization. A short ORL often represents a narrow query for which a displacement matters more, whereas a long ORL often represents a broad query with many acceptable answers. Exp7 and Exp8 use the same candidate and denominator and differ only in ORL size, so they isolate the size effect: the rise in \(M_2\) comes from \(\sqrt{|R_q|}\) weakening the penalty. \RDQmi{} stays flat at 0.7619 because \(M_1\) has no size term. MAP decreases and \NDCGlinear{} drops slightly. Exp6 differs additionally in the number of retrieved reference items and in \(N_k\), so its gap from Exp7 is not due to ORL size alone.

\emph{Output-position importance (Exp9--10).}\label{ex:ce3}
Both experiments use the same ORL and prediction, where the prediction places an item outside the ORL at position 2. Only the output-position weights differ. Exp9 uses the default vertical-search weights, and Exp10 uses $\mu^\star=[1,0.2,0.2,0.1,0.1]$, modeling a layout where the first position gets most of the attention. Because the prediction is identical, MAP, RBP, the NDCG variants, and \CMP{} give the same score in both rows, and the rank correlations are omitted. RDQ changes because it includes the output-position weight $\mu_i$. In Exp10 the item outside the ORL at position 2 has less impact, so the score rises from 0.7429 to 0.8667. The three RDQ variants coincide because every retrieved reference item sits at its target rank, so the penalty function does not matter and only the weights change.

\emph{Reversed order and tied relevance (Exp11--13).}\label{ex:ce4}
Here the ORL is $\{d_1{:}1,d_2{:}2\}$. Exp11 is the ideal order and Exp12 reverses it. In Exp12, $\tau$ and $\rho$ assign $-1$, while the effectiveness metrics (\NDCGrecip{}, \NDCGlinear{}, \CMP{}, and the RDQ variants) instead assign graded partial credit. MAP, RBP, and \NDCGb{} stay unchanged. The order-aware effectiveness metrics thus differ in penalty severity. Exp13 instead ties both items at rank 1 with $\{d_1{:}1,d_2{:}1\}$. Standard $\tau$ and $\rho$ are then undefined because the reference-rank vector is constant. \NDCGrecip{}, \NDCGlinear{}, \CMP{}, and RDQ all assign full credit because the two items belong to the same tier.

Taken together, the examples separate the metrics by what they can and cannot see. Metrics built on binary relevance (MAP, RBP, \NDCGb{}) give the same score whenever the same reference items are returned, no matter their order. Rank correlations are order-sensitive but require a common-item transformation when the lists differ and are undefined for an all-tied reference. \NDCGrecip{}, \NDCGlinear{}, and \CMP{} fall in between, handling order and ties but not output-position weights or ORL size. Among the metrics compared, only the RDQ \(M_2\) variants respond to every property tested here, with $\alpha$ controlling how strongly a displacement is penalized. 

\vspace{-.3em}
\subsection{Experiments on the POI Dataset}\label{sec:internal}

This section evaluates RDQ on a POI dataset whose ORL sizes vary widely across queries. For example, the ORL for ``restaurants in Tacoma'' contains 47 reference POIs, while ``Hotel Indigo Lower East Side'' has one\cite{shah2026omniqum}. We analyze pairwise discriminative power, system-ranking stability under reduced query budgets, and parameter sensitivity.

\vspace{-.3em}
\subsubsection{Experiment Setup}

\vspace{-.3em}
\paragraph{Dataset.}

The document corpus consists of POI records drawn from the Overture Maps open database~\cite{overturemaps}. We use a stratified sample of 5,000 POISS queries~\cite{poidataset2026} covering different ORL sizes. POISS derives its labels from an external ranking signal and uses generative AI for labeling\cite{gsl}. We convert the POISS labels into query-specific ORLs as described in Section~\ref{sec:orl_rank}.

The sample contains 100 queries with $\leq 3$ reference items and 4{,}900 queries with more than three. This covers both short-ORL queries, which often correspond to factual intents, and broader queries with many acceptable answers.

\vspace{-.3em}

\paragraph{Metric configurations.}\label{sec:new_baseline_metrics} 
The main metric configurations and their default settings are listed in Section~\ref{sec:poi-common-setup}. For the POI analysis, we expand RBP to $p\in\{0.5,0.8,0.9\}$ and CMP to $p\in\{0.8,0.9,0.95\}$, giving 13 metric configurations in total.
We do not use Kendall's $\tau$ or Spearman's $\rho$ as retrieval/ranking metrics because candidate and reference lists may contain different items. Kendall's $\tau$ is used separately to compare complete system rankings in the stability and correlation analyses.

\vspace{-.3em}
\paragraph{System rankings.}\label{sec:sys_12} 
We evaluate 12 system configurations formed by combining four cross-encoder models based on the H2CE architecture~\cite{bai2026h2ce} with three inference feature settings, as shown in Table~\ref{tab:own_systems}. These 12 configurations define 66 pairwise system comparisons.

Using all 5{,}000 queries, we test every system pair under each of the 13 metric configurations with a two-sided randomization test (1{,}000 permutations; $\alpha=0.05$ for each test). For 46 of the 66 pairs (70\%), the test rejects the null hypothesis of no system difference under all 13 metric configurations. It rejects the null under at least seven configurations for 60 pairs (91\%), and under at least one configuration for 64 pairs (97\%); for the remaining two pairs, it rejects under none. In other words, the systems in each of these 46 pairs are statistically distinguishable according to every evaluated metric configuration.

The analyses below compare the power of these tests under reduced query budgets and the stability of the resulting 12-system rankings.

\begin{table}[t]
  \centering
  \caption{System configurations: each of the 4 models is evaluated under all three feature settings, yielding 12 rankings.}
  \vspace{-.3em}
  \label{tab:own_systems}
\hspace{-2em}
  \begin{minipage}[t]{0.46\linewidth}
  \centering
  \footnotesize
  \textbf{(a) Model variants.} \\[4pt]
  \begin{tabular}{@{}ll@{}}
  \toprule
  \textbf{Model} & \textbf{Training variation} \\
  \midrule
  V72 & \hspace{-1.5em}Baseline \\
  V74 & \hspace{-1.5em}Alternative sampling \\
  V75 & \hspace{-1.5em}More negatives per example \\
  V81 & \hspace{-1.5em}Modified distance feature \\
  \bottomrule
  \end{tabular}
  \end{minipage}
  \vspace{.8em}
  \begin{minipage}[t]{0.46\linewidth}
  \centering
  \footnotesize
  \textbf{(b) Inference feature settings.} \\[4pt]
  \begin{tabular}{@{}ll@{}}
  \toprule
  \textbf{Setting} & \textbf{Features} \\
  \midrule
  MP3 & \hspace{-1.5em}Name, address, all categories \\
  MP4 & \hspace{-1.5em}Name, addr., distance, top-1 category \\
  MP5 & \hspace{-1.5em}Name, addr., distance, all categories \\
  \\
  \bottomrule
  \end{tabular}
    \end{minipage}
\vspace{-1em}
\end{table}


\subsubsection{Discriminative Power Analysis}\label{sec:disc-power}
\vspace{-.6em}
\paragraph{All queries.}
The full dataset defines all 66 pairwise comparisons, but it does not reveal how many queries are needed to reach reliable conclusions. Higher empirical power means that, for the score differences induced by a metric on the observed system pairs, fewer sampled queries are required to reject the null hypothesis. It indicates greater statistical sensitivity to the supplied reference signal, but does not establish that the detected distinctions better reflect user preferences. We therefore measure how many queries each metric needs to reliably detect system differences.

We use two-sided randomization tests~\cite{smucker2007comparison} with 1{,}000 permutations~\cite{Marozzi_2004} at \(\alpha=0.05\) for all pairwise system comparisons. For each pair of systems, the null hypothesis is that the two systems produce identical per-query metric score distributions. We estimate discriminative power by Monte Carlo simulation: for each system pair and sample size \(n \in \{25, 50, 75, 100, \ldots, 500, 750, 1000, 1500, 2000\}\), we draw 300 random subsamples of \(n\) queries, apply the randomization test to each, and record the rejection proportion as estimated power. This yields Monte Carlo SE at most \(0.029\)~\cite{Burton2006simulation,Morris2019simulation}.

From these estimates, we report two summary statistics. First, at a fixed sample size of \(n=100\), we compute the \nth{10}, \nth{50}, and \nth{90} percentiles of power across all system pairs, denoted \(p_{10}\), \(p_{50}\), and \(p_{90}\). The lower percentile reflects difficult system pairs that metrics struggle to distinguish, while the upper percentile reflects easier pairs. Second, we report \(N_{80}\), the minimum sample size at which estimated power reaches \(80\%\). If a system pair does not reach \(80\%\) power within the simulated range, we set \(N_{80}=\infty\). We report the median \(N_{80}\) across all system pairs; lower values indicate that a metric requires fewer queries to reliably detect system differences.

Table~\ref{tab:disc-power} reports discriminative power across 66 system pairs. Median power@100 is 0.353 for \RDQma@5, 0.292 for \RDQmb@5, and 0.290 for \RDQmi@5. The inverse- and linear-conversion NDCG variants reach median power of 0.233 and 0.213, respectively. Across the tested settings, RBP is strongest at $p=0.9$ with median power@100 of 0.287, while CMP is strongest at $p=0.95$ with 0.217. \RDQma@5 remains highest at 0.353, 23\% above RBP(0.9), and has a lower median $N_{80}$ (350 versus 450).

\begin{table}[t]
    \caption{Discriminative power across 66 system pairs: percentiles of power@100 and median $N_{80}$ over all queries, and median power@100 within each reference-list size bucket. Metrics are sorted by all-query $p_{50}$. Largest power values and smallest $N_{80}$ are bold.}
    \label{tab:disc-power}
    \centering
    \small
    \setlength{\tabcolsep}{3pt}
    \begin{tabular}{l cccc @{\hspace{1em}} cccc}
    \toprule
     & \multicolumn{4}{c}{All queries} & \multicolumn{4}{c}{By reference-list size} \\
    \cmidrule(lr){2-5} \cmidrule(lr){6-9}
    \textbf{Metric} & $p_{10}$ & $p_{50}$ & $p_{90}$ & $N_{80}$ &
    $\leq 3$ & $(3,10]$ & $(10,20]$ & $>20$ \\
    \midrule
    \footnotesize{\RDQma@5}       & 0.062          & \textbf{0.353} & 0.798          & \textbf{350} & 0.000          & 0.105          & \textbf{0.258} & \textbf{0.592} \\
    \footnotesize{\RDQmb@5}       & 0.058          & 0.292          & 0.743          & 400           & \textbf{0.073} & 0.103          & 0.245          & 0.502 \\
    \footnotesize{\RDQmi@5}       & 0.053          & 0.290          & \textbf{0.823} & 375           & 0.013          & \textbf{0.117} & 0.243          & 0.575 \\
    \footnotesize{RBP(0.9)@5}     & \textbf{0.067} & 0.287          & 0.710          & 450           & 0.000          & 0.105          & 0.232          & 0.460 \\
    \footnotesize{RBP(0.8)@5}     & 0.060          & 0.243          & 0.680          & 500           & 0.000          & 0.082          & 0.200          & 0.452 \\
    \footnotesize{\NDCGrecip@5}  & 0.048          & 0.233          & 0.707          & 625           & 0.000          & 0.107          & 0.182          & 0.427 \\
    \footnotesize{\NDCGb@5}      & 0.058          & 0.230          & 0.680          & 475           & 0.000          & 0.080          & 0.198          & 0.427 \\
    \footnotesize{\CMP(0.95)@5}  & 0.043          & 0.217          & 0.683          & 750           & 0.000          & 0.108          & 0.175          & 0.373 \\
    \footnotesize{\NDCGlinear@5} & 0.055          & 0.213          & 0.687          & 750           & 0.000          & 0.093          & 0.162          & 0.362 \\
    \footnotesize{\CMP(0.9)@5}   & 0.055          & 0.203          & 0.638          & 500           & 0.000          & 0.098          & 0.173          & 0.397 \\
    \footnotesize{\CMP(0.8)@5}   & 0.050          & 0.200          & 0.613          & 750           & 0.000          & 0.083          & 0.167          & 0.352 \\
    \footnotesize{MAP@5}          & 0.053          & 0.177          & 0.462          & 1000          & 0.000          & 0.080          & 0.203          & 0.403 \\
    \footnotesize{RBP(0.5)@5}     & 0.055          & 0.165          & 0.525          & 750           & 0.000          & 0.053          & 0.137          & 0.295 \\
    \bottomrule
    \end{tabular}
\vspace{-2em}
\end{table}

\paragraph{By reference-list size.}
The aggregate results treat all queries equally. However, queries with few reference items represent exact-answer retrieval where each misplacement is critical, while queries with many reference items represent broader ranking tasks. To examine how discriminative power changes with reference-list size, we partition queries into four groups: $\leq 3$, $(3,10]$, $(10,20]$, and $>20$, containing 100, 762, 2{,}397, and 1{,}741 queries, respectively. We apply the same randomization test within each group (Table~\ref{tab:disc-power}).

Within the $\leq 3$ bucket, RDQ is the only metric family with nonzero median power: \RDQmb@5 reaches 0.073 and \RDQmi@5 reaches 0.013, while \RDQma@5 and all non-RDQ configurations are zero. This bucket contains exactly 100 queries and is evaluated once rather than by subsampling, so the reported values are point estimates without error bars. Power increases with reference-list size for every configuration. In the $>20$ bucket, RBP(0.9)@5, the strongest non-RDQ configuration, reaches 0.460, while \RDQma@5, \RDQmi@5, and \RDQmb@5 reach 0.592, 0.575, and 0.502, respectively. The gap between \RDQma@5 and RBP(0.9)@5 increases from 0.066 to 0.132 in this bucket. Thus, RDQ's median-power advantage is larger for queries with long reference lists in this dataset.

\vspace{-.3em}
\subsubsection{System Ranking Stability}\label{sec:ranking-stability}

Discriminative power measures whether a metric configuration can distinguish individual system pairs. We also evaluate whether it produces stable overall system rankings under reduced query budgets. Following the subsampling approach of \citet{urbano2013reliability}, we use the full set of 5{,}000 queries and 12 systems to compute a full-query reference ranking for each of the 13 metric configurations. Systems are ranked by their mean per-query score over all queries. This analysis compares each configuration with its own full-query system ranking and therefore does not assume that different configurations induce the same reference ordering.

To measure stability at reduced sample sizes, we draw random subsamples of \(n\) queries (without replacement), compute mean system scores on each subsample, rank the 12 systems, and measure Kendall's \(\tau\) between the subsample system ranking and the full-query reference ranking. Here, \(\tau\) measures agreement between system orderings, not retrieval quality. We repeat this process 300 times for each subsample size and report the mean \(\tau\).

We summarize with five statistics: \(\tau_{25}\), \(\tau_{100}\), and \(\tau_{200}\) (mean \(\tau\) at those sample sizes), \AUCt{} (normalized area under the \(\tau\)-vs.-\(n\) curve), and \(n@\tau{\geq}0.8\) (minimum sample size to reach \(\tau \geq 0.8\)). This analysis complements Section~\ref{sec:disc-power}: discriminative power tests whether individual system pairs are distinguishable, while ranking stability tests whether the complete system ordering is preserved under reduced query budgets.

\begin{table}[t]
   \centering
   \caption{System-ranking stability under query subsampling (300 repetitions per sample size). Metrics are sorted by \AUCt{} descending.}
   \label{tab:ranking-stability}
   \small
   \begin{tabular}{@{}lccccc@{}}
   \toprule
   \textbf{Metric} & \(\tau_{25}\) & \(\tau_{100}\) & \(\tau_{200}\) & \AUCt{} & \(n@\tau{\geq}0.8\) \\
   \midrule
   \RDQma@5        & \textbf{0.483} & \textbf{0.711} & \textbf{0.811} & \textbf{0.904} & \textbf{200} \\
   \RDQmb@5        & 0.475 & 0.675 & 0.783 & 0.881 & 250 \\
   RBP(0.9)@5      & 0.444 & 0.696 & 0.774 & 0.874 & 250 \\
   RBP(0.8)@5      & 0.454 & 0.677 & 0.776 & 0.870 & 250 \\
   \RDQmi@5        & 0.443 & 0.675 & 0.758 & 0.865 & 300 \\
   \NDCGb@5        & 0.423 & 0.665 & 0.760 & 0.864 & 300 \\
   \CMP(0.95)@5    & 0.397 & 0.626 & 0.741 & 0.851 & 350 \\
   \CMP(0.9)@5     & 0.412 & 0.627 & 0.720 & 0.849 & 400 \\
   \CMP(0.8)@5     & 0.366 & 0.624 & 0.705 & 0.846 & 450 \\
   \NDCGlinear@5   & 0.387 & 0.622 & 0.731 & 0.846 & 400 \\
   RBP(0.5)@5      & 0.365 & 0.600 & 0.717 & 0.839 & 400 \\
   \NDCGrecip@5    & 0.437 & 0.630 & 0.739 & 0.839 & 400 \\
   MAP@5            & 0.343 & 0.531 & 0.678 & 0.818 & 750 \\
   \bottomrule
   \end{tabular}
\end{table}

Table~\ref{tab:ranking-stability} reports system-ranking stability under query subsampling. \RDQma@5 has the highest agreement at each reported sample size, with \AUCt{} $=0.904$, and reaches $\tau\geq0.8$ with 200 queries. \RDQmb@5 has \AUCt{} $=0.881$ and reaches the threshold with 250 queries. Across the tested RBP settings, RBP(0.9)@5 has the highest \AUCt{} at 0.874; both RBP(0.9)@5 and RBP(0.8)@5 reach the threshold with 250 queries, while RBP(0.5)@5 requires 400. \RDQmi@5 has \AUCt{} $=0.865$ and requires 300 queries, so the two $M_2$ variants, not $M_1$, drive RDQ's stability advantage.

Among the remaining baselines, \NDCGb@5 has \AUCt{} $=0.864$ and reaches the threshold with 300 queries. \CMP(0.95)@5 is the strongest CMP setting at 0.851 and 350 queries, while the linear- and inverse-conversion NDCG variants reach 0.846 and 0.839, each requiring 400 queries. MAP@5 has the lowest \AUCt{} at 0.818 and requires 750 queries. Thus, \RDQma@5 is the only evaluated configuration to reach mean $\tau\geq0.8$ with 200 queries; the strongest RBP setting by \AUCt{} reaches it with 250.

\vspace{-.3em}
\subsubsection{Sensitivity to \(M_2\) Parameters}\label{sec:sensitivity}
We focus on \(M_2\) because $\alpha$ and $\lambda$ are specific to this penalty. \(M_1\) has no corresponding parameters. We evaluate all 25 combinations of $\alpha \in \{0.5, 1, 2, 4, 8\}$ and $\lambda \in \{0, 0.05, 0.1, 0.2, 0.4\}$ while keeping the default UI position weights $\mu$ fixed. We use the same testing protocol as Section~\ref{sec:disc-power}.

\begin{figure}[t]
    \centering
    \includegraphics[width=0.7\columnwidth]{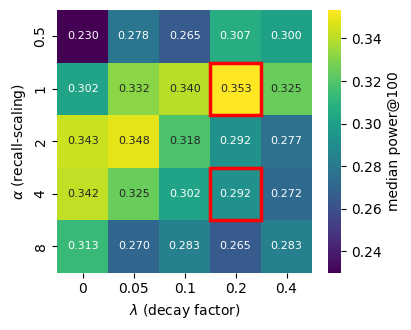}
    \vspace{-1.5em}
        \captionof{figure}{Median power@100 of RDQ ($M_2$, $k{=}5$) across all 25 combinations of $\alpha$ and $\lambda$. The red boxes mark the two configurations reported in the main experiments: $(\alpha{=}1,\lambda{=}0.2)$ and $(\alpha{=}4,\lambda{=}0.2)$. Sixteen configurations exceed RBP(0.9)@5, the strongest tested non-RDQ configuration, at 0.287.}
     \vspace{-1em}
    \label{fig:sensitity-analysis}
\end{figure}

As shown in Figure~\ref{fig:sensitity-analysis}, median power@100 across the grid ranges from 0.230 to 0.353, with a median of 0.302. Sixteen of the 25 configurations exceed RBP(0.9)@5 at 0.287, the strongest tested non-RDQ configuration, and 24 exceed the default RBP(0.8)@5 at 0.243. Both parameters increase displacement tolerance through separate factors in the penalty denominator: $\alpha$ globally and $1+\lambda(r-1)$ according to reference rank. 
The highest estimates occur at intermediate values of $\alpha$ from 1 to 4. Power peaks at 0.353 for $(\alpha=1,\lambda=0.2)$ and decreases toward both the strictest settings and the most lenient $\alpha=8$ row.
The strictest setting, $(\alpha{=}0.5,\lambda{=}0)$, has the lowest power at 0.230 and is below both RBP settings. At $\alpha{=}8$, power ranges from 0.265 to 0.313: all five settings exceed RBP(0.8), but only $\lambda=0$ exceeds RBP(0.9). The two configurations reported in the main experiments, $(1,0.2)$ and $(4,0.2)$, both exceed RBP(0.9). Thus, higher median power than tuned RBP is not limited to one setting, but it does not hold throughout the parameter grid.

\vspace{-.3em}
\subsection{Experiments on Standard IR Benchmarks}\label{sec:ir-benchmarks}

Finally, we evaluate RDQ on standard IR benchmarks with graded relevance judgments to measure its behavior on tie-heavy rankings derived from graded labels.

\begin{table}[t]
  \centering
  \vspace{-.5em}
  \caption{Discriminative power on pooled TREC-DL 2019--2022 system pairs.
  \(p_{10}\), \(p_{50}\), and \(p_{90}\) denote percentiles of statistical
  power. Metrics are sorted by \(p_{50}@25\) descending.}
  \label{tab:trec}
  \small
  \begin{tabular}{@{}lcccccc@{}}
  \toprule
  & \multicolumn{3}{c}{\(n = 25\)} & \multicolumn{3}{c}{\(n = 100\)} \\
  \cmidrule(lr){2-4} \cmidrule(lr){5-7}
  \textbf{Metric} & \(p_{10}\) & \(p_{50}\) & \(p_{90}\) & \(p_{10}\) & \(p_{50}\) & \(p_{90}\) \\
  \midrule
  \RDQmi@10        & \textbf{0.060} & \textbf{0.108} & 0.232 & 0.052 & 0.367 & \textbf{0.870} \\
  MAP@10           & 0.042 & 0.107 & 0.193 & 0.035 & 0.338 & 0.468 \\
  NDCG@10          & 0.040 & 0.102 & \textbf{0.265} & 0.047 & \textbf{0.437} & 0.868 \\
  \RDQma@10        & 0.038 & 0.095 & 0.168 & 0.050 & 0.300 & 0.633 \\
  \NDCGb@10        & 0.033 & 0.085 & 0.165 & 0.038 & 0.243 & 0.652 \\
  CMP(0.95)@10     & 0.052 & 0.072 & 0.155 & 0.022 & 0.245 & 0.643 \\
  RBP(0.9)@10      & 0.038 & 0.068 & 0.205 & 0.030 & 0.247 & 0.583 \\
  RBP(0.8)@10      & 0.038 & 0.068 & 0.145 & \textbf{0.058} & 0.218 & 0.658 \\
  \RDQmb@10        & 0.035 & 0.067 & 0.130 & 0.042 & 0.268 & 0.595 \\
  RBP(0.5)@10      & 0.033 & 0.067 & 0.108 & 0.017 & 0.185 & 0.370 \\
  CMP(0.8)@10      & 0.040 & 0.060 & 0.122 & 0.018 & 0.157 & 0.450 \\
  CMP(0.9)@10      & 0.035 & 0.053 & 0.162 & 0.017 & 0.170 & 0.565 \\
  \bottomrule
  \end{tabular}
  \vspace{-1em}
\end{table}

\subsubsection{Experiment Setup}
\vspace{-.3em}
\paragraph{Dataset.}

We use four TREC Deep Learning track datasets (2019--2022)~\cite{Craswell2019TrecDl,Craswell2020TrecDl,craswell2025overviewtrec2021deep,craswell2025overviewtrec2022deep}, each with relevance judgments on a four-point scale ($0$ = not relevant to $3$ = highly relevant). 
All four years are evaluated against the MS~MARCO~v2 document corpus. For 2019--2020, we use the official v1-to-v2 migrated qrels distributed through \texttt{ir\_datasets}~\cite{irdatasets}\footnote{\label{fn:trec-dl-judged} \texttt{msmarco-document-v2/trec-dl-\{year\}/judged}}, which omit unmatched v1 documents.
The four datasets contain 43, 45, 57, and 76 judged queries respectively. We pool them into a single evaluation set of 221 queries, where each query retains its original relevance judgments. Pooling increases the number of queries available for subsampling, yielding more stable power estimates than per-dataset analysis with only 43--76 queries each. The 221 pooled queries have an average of 498 relevant documents per query. Because these documents span fewer than three positive grades per query on average, they form large tied tiers, with a median size of 21 and a mean size of 177.

\vspace{-.3em}
\paragraph{Retrieval and reranking}
We retrieve the top 100 documents per query from the MS MARCO v2 document
collection~\cite{Bajaj2016Msmarco} using Lucene BM25~\cite{apachelucene}.
We then rerank these candidates with four pretrained cross-encoder models:
\texttt{ms-marco-MiniLM-L6-v2} and \texttt{ms-marco-MiniLM-L12-v2}
\cite{reimers2019sentence,wang2020minilm}, and
\texttt{bge-reranker-base} and \texttt{bge-reranker-large}
\cite{bge_embedding}. Each model produces a distinct ranking per query,
yielding four systems. The models range from 22.7 million to 0.6 billion parameters and come from two independent research groups. Their different architectures and training methods provide varied system outputs for metric comparison.

\vspace{-.3em}
\paragraph{Metrics}
All metrics are evaluated at cutoff \(k=10\). We report native NDCG using the original grades; MAP, RBP, and \NDCGb{} use binary relevance. RDQ and \CMP{} use the ORL tiers derived below. Similar to the POI dataset experiment, we evaluate RBP with $p\in\{0.5,0.8,0.9\}$ and \CMP{} with $p\in\{0.8,0.9,0.95\}$. We report power at both \(n=25\) and \(n=100\): \(n=25\) reflects the scale of individual TREC-DL datasets, while \(n=100\) enables comparison with the POI analysis in Section~\ref{sec:disc-power}.

RDQ consumes the item-to-ORL-rank mapping defined in Section~\ref{sec:orl_rank}. To construct it from graded relevance judgments, we keep only documents with positive grades. Each distinct grade becomes a tier: the highest grade maps to ORL rank 1, the next to rank 2, and so on. For example, grades ${d_1{:}3, d_2{:}1, d_3{:}1, d_4{:}0}$ produce ${d_1{:}1, d_2{:}2, d_3{:}2}$, and $d_4$ is excluded.
  
\vspace{-.3em}
\subsubsection{Discriminative Power on Standard Benchmarks}

We apply the same Monte Carlo power simulation as Section~\ref{sec:disc-power} to the pooled TREC-DL dataset. Table~\ref{tab:trec} reports discriminative power across six system pairs. At $n=25$, median power is 0.108 for \RDQmi@10, 0.107 for MAP@10, and 0.102 for native NDCG@10; these estimates are close. At $n=100$, median power is 0.437 for native NDCG@10 and 0.367 for \RDQmi@10, while their $p_{90}$ values are 0.868 and 0.870. Across the tested persistence settings, \CMP{} is strongest at $p=0.95$, reaching median power 0.082 at $n=25$ and 0.237 at $n=100$, and RBP is strongest at $p=0.9$, reaching 0.068 and 0.247; at $n=100$, no RBP or \CMP{} setting exceeds any RDQ variant or native NDCG.

\vspace{-.3em}
\subsubsection{Agreement with Established Metrics}\label{sec:ir-sys-corr}
\begin{figure}[t]
    \centering
    \includegraphics[width=\columnwidth]{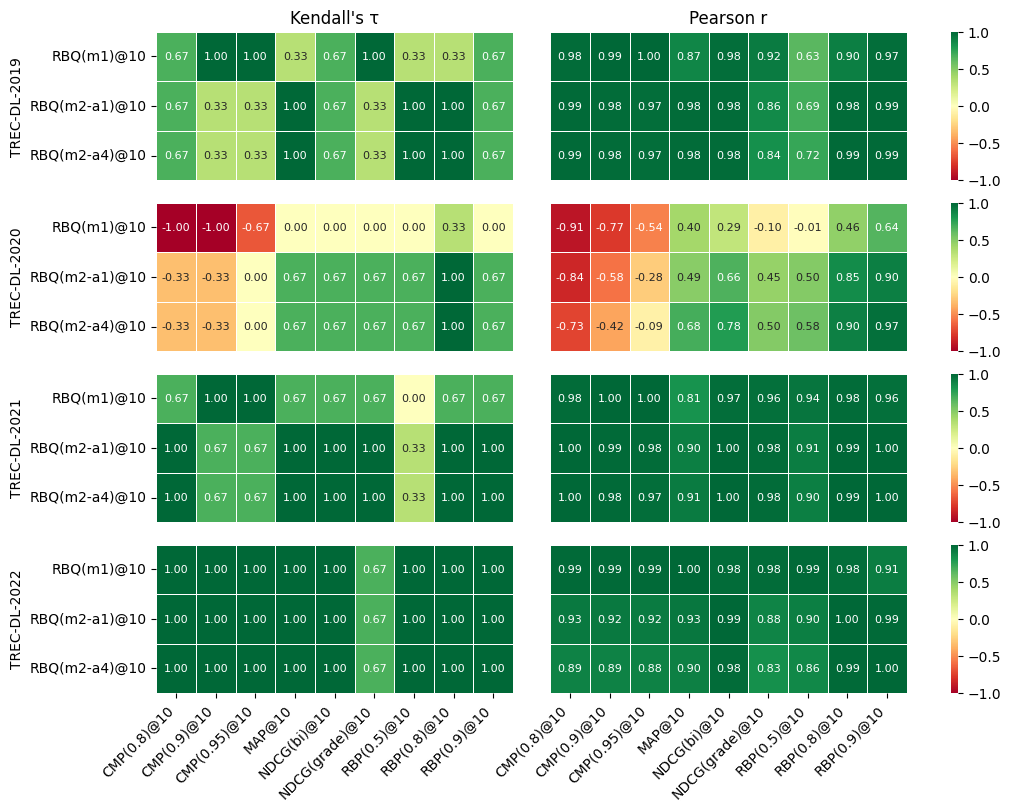}
        \caption{Agreement between RDQ variants and nine baseline metric configurations across four TREC-DL datasets.}
    \Description{Heatmaps of Kendall $\tau$ and Pearson $r$ between three RDQ variants and MAP, binary NDCG, native NDCG, CMP at three persistence settings, and RBP at three persistence settings for TREC-DL 2019--2022. Agreement is strongest in 2021--2022 and weakest for RDQ m1 in 2020.}
    \label{fig:sys-correlation}
\end{figure}

We next explore whether RDQ orders systems similarly to established metrics, and where it diverges. For each TREC-DL year, we compute Kendall’s $\tau$ and Pearson’s $r$ between each RDQ variant and each baseline over the four system scores (Figure~\ref{fig:sys-correlation}). We report correlations per year rather than pooled to show the year-specific disagreements.

Agreement is generally high in 2021 and 2022. Pearson's $r$ is between 0.81 and 1.0 for every pair, and Kendall's $\tau$ is at least 0.67 except comparisons with RBP(0.5) in 2021 (ranging from 0.00 to 0.33). Agreement is more mixed in 2019. Pearson's $r$ ranges from 0.84 to 1.0 for most baselines but drops to 0.63--0.72 against RBP(0.5). Kendall's $\tau$ ranges from 0.33 to 1.0, indicating differences among some four-system orderings.
These results provide a descriptive consistency check for the four tested systems across the four TREC-DL datasets. RDQ generally agrees with established metrics but produces different system orderings in some cases.

TREC-DL 2020 shows the weakest agreement, and the disagreement centers on CMP. \RDQmi{} ranks the four systems in the exact opposite order of CMP(0.8) and CMP(0.9) ($\tau=-1.0$) and shows little agreement with the other baselines ($\tau \leq 0.33$). The $M_2$ variants agree with the non-CMP baselines ($\tau \geq 0.67$) but disagree with all three CMP settings ($\tau$ between $-0.33$ and 0). Pearson's $r$ shows the same pattern. With only four systems, these correlations are coarse: a single rank swap can change them. We therefore read them qualitatively, as showing where RDQ agrees with established metrics and where it diverges, not as precise estimates.

\vspace{-.3em}
\subsection{Discussion}\label{sec:discussion}

The relative performance of the three variants depends on the structure of the reference. POI ORLs assign nearly unique ranks, whereas grade-derived TREC ORLs contain a few large tiers. $M_2$ measures displacement in ORL-rank units but scales its penalty by $\sqrt{|R_q|}$, which counts items, so large tiers weaken the penalty. $M_1$ compares ORL ranks as a ratio and does not depend on list size. The results are consistent with this mechanism. On POI, the $M_2$ variants lead: \RDQma{} has the highest aggregate median, and \RDQmb{} reaches 0.073 in the $\leq3$ bucket, where every other metric is at or below 0.013. On TREC-DL, $M_1$ has the largest RDQ median. The TREC result is consistent with this mechanism, but the four-system panel is small to support a precise effect estimate. As practical guidance: use $M_2$ with $\alpha\in[1,4]$ and $\lambda=0.2$ by default, use a smaller $\alpha$ when exact placement matters most, and consider $M_1$ when the reference contains large tied tiers. More broadly, When calibrated graded judgments already exist, native NDCG remains a strong choice. RDQ instead targets ordinal references, such as those constructed from pairwise or listwise preferences.

These results also show how the reference representation affects performance. Pairwise or listwise judgments can provide an ordered reference without requiring globally calibrated grade boundaries~\cite{carterette2008here,scholer2013effect}. A common workaround is to map ranks to relevance values and apply a graded metric. On the same POI reference signal, \NDCGrecip{} reaches median power 0.233 and \NDCGlinear{} reaches 0.213, compared with 0.353 for \RDQma{}. The two mappings therefore support the same median-power conclusion, although they do not exhaust all possible relevance mappings or establish lower annotation cost. Part of RDQ's resolution advantage may also be mechanical rather than semantic. RDQ’s continuous penalty can produce fewer tied per-query scores than MAP@5. Fewer score ties can make small differences between systems easier for the randomization test to detect.

Several limitations should be noted. (1) RDQ requires application choices for the penalty function and UI weights; $M_2$ additionally uses $\alpha$ and $\lambda$. The penalty forms are chosen for their intended properties (Section~\ref{sec:penalty}) rather than derived from a model of user behavior. UI importance weights are set manually, and learning them from interaction data is left to future work. We test $M_1$ and $M_2$ and evaluate 25 $M_2$ parameter configurations (Section~\ref{sec:sensitivity}).
(2) Cross-format comparisons require conversion choices in both directions. RDQ on TREC-DL needs a grade-to-rank conversion, which discards the spacing between grades. NDCG on POI needs a rank-to-grade conversion, which requires gain values that the ordinal judgments do not provide. The evaluated conversions and persistence settings do not exhaust all plausible choices.
(3) The TREC-DL evaluation uses only four systems (six pairs), producing noisy estimates. 
(4) The $\leq 3$ bucket contains exactly 100 queries, so power@100 there reflects a single evaluation rather than a subsampling estimate. 
(5) Discriminative power and ranking stability are internal statistical properties. They measure how sensitively and consistently a metric separates systems, not whether the differences they reward correspond to user-perceived quality. Validating RDQ against human preference judgments is left to future work.
(6) The POI ORLs come from a GenAI-assisted labeling pipeline and act as a silver-standard proxy rather than human-annotated ground truth. A system that genuinely surpasses this reference may be under-credited, since the reference pipeline becomes a ceiling on measurable performance~\cite{Soboroff_2025}, and label noise may affect metrics differently.
(7) RDQ assumes the ORL is sufficiently complete (Section~\ref{sec:orl_rank}); with an incomplete reference, valid but unlisted items receive zero credit, penalizing systems that retrieve them.

The POI experiment uses a 5,000-query subsample of POISS~\cite{poidataset2026}, selected to bound the computational cost of the power simulations. At submission time, POISS is not yet publicly available but is scheduled for release. The controlled examples provide reproducible checks of RDQ's behavior, and the TREC-DL experiments provide a complementary public-data evaluation using human relevance judgments.

Future work will evaluate agreement with human preferences, annotation cost, robustness across larger system pools, and data-driven selection of application-specific weights.

\vspace{-.3em}
\section{Conclusion}\label{sec:conclusion}

We propose Rank-Deviation Quality (RDQ), an effectiveness metric for multi-answer retrieval and ranking that evaluates system outputs against an ordered reference list, without requiring graded gain values. RDQ supports tied tiers, query-varying reference sizes, and application-specific position weights. 
On the 12-system POI panel, \RDQma{} has the highest median power@100 (0.353 versus 0.287 for RBP(0.9), the strongest tested non-RDQ configuration). It reaches mean $\tau \geq 0.8$ agreement with its own full-query ordering at 200 queries versus 250 for RBP(0.9); 16 of the 25 $M_2$ configurations exceed RBP(0.9) in median power. On the four-system TREC-DL panel, \RDQmi{} is comparable to native graded NDCG at $n=25$ (0.108 versus 0.102), while NDCG is higher at $n=100$ (0.437 versus 0.367).
These experiments characterize RDQ's statistical sensitivity and subsampling consistency with respect to the supplied ordinal references. Native NDCG remains a strong choice when calibrated graded judgments are available. RBP and other persistence-based metrics are better suited when evaluation should model user browsing behavior. RDQ targets the complementary setting of ordinal references.





\vspace{-.3em}
\section*{Ethical Considerations}
As an evaluation metric, RDQ can affect systems and decisions through the rankings it favors. RDQ inherits any errors or biases in its reference. For example, the POI ORLs use GenAI-assisted, silver-standard labels. For high-stakes decisions, metric scores should be paired with human evaluation. No personal or identifying information is used at any stage. The ordinal judgments RDQ consumes describe items, not users.
The authors used AI assistants to draft and polish text from author-provided outlines and feedback, format LaTeX, and verify statistical simulation code. All research contributions and technical content are the authors' own work.

\FloatBarrier
\bibliographystyle{ACM-Reference-Format}
\bibliography{references}

\end{document}